# Superconducting Dome in Nd$_{1-x}$Sr$_x$NiO$_2$ Infinite Layer Films

Danfeng Li[1,2,\*,†], Bai Yang Wang[1,3,\*,‡], Kyuho Lee[1,3], Shannon P. Harvey[1,2], Motoki Osada[1,4], Berit H. Goodge[5], Lena F. Kourkoutis[5,6] and Harold Y. Hwang[1,2,§]

[1]*Stanford Institute for Materials and Energy Sciences, SLAC National Accelerator Laboratory, Menlo Park, California 94025, USA*
[2]*Department of Applied Physics, Stanford University, Stanford, California 94305, USA*
[3]*Department of Physics, Stanford University, Stanford, California 94305, USA*
[4]*Department of Materials Science and Engineering, Stanford University, Stanford, California 94305, USA*
[5]*School of Applied and Engineering Physics, Cornell University, Ithaca, New York 14853, USA*
[6]*Kavli Institute at Cornell for Nanoscale Science, Ithaca, New York 14853, USA*

ABSTRACT

We report the phase diagram of Nd$_{1-x}$Sr$_x$NiO$_2$ infinite layer thin films grown on SrTiO$_3$. A superconducting dome spanning $0.125 < x < 0.25$ is found, remarkably similar to cuprates, albeit over a narrower doping window. However, while cuprate superconductivity is bounded by an insulator for underdoping and a metal for overdoping, here we observe weakly insulating behavior on either side of the dome. Furthermore, the normal state Hall coefficient is always small and proximate to a continuous zero crossing in doping and in temperature, in contrast to the ~1/$x$ dependence observed for cuprates. This suggests the presence of both electron- and hole-like bands, consistent with band structure calculations.

A common feature of many unconventional superconductors is a doping-dependent superconducting dome, often in proximity to a competing ordered phase [1-3]. The generic structure of their phase diagrams raises questions regarding the contribution to pairing by electron-phonon coupling, for which such a doping dependence is not *a priori* obvious, and suggests the possibility of coupling to fluctuations of the competing order. The recent finding of superconductivity [4] in $Nd_{0.8}Sr_{0.2}NiO_2$ was motivated by long-standing consideration of possible analogies, and distinctions, between infinite layer nickelates and cuprates [5,6]. The undoped parent compound for both of these materials share the same nominal transition metal $3d^9$ configuration and crystal structure. This specific composition was identified, perhaps serendipitously, by noting the approximate composition of maximum conductivity in $(La,Sr)NiO_2$, with the aim of increasing the electronic bandwidth using the smaller Nd in place of La [4]. Of course, the fact that this level of Sr substitution was near optimal for cuprate superconductivity was also encouraging [7]. However, unlike the antiferromagnetic insulator observed in undoped cuprates, thus far there is no evidence for an ordered ground state in the undoped infinite layer nickelates [8,9], although recent theoretical work suggests that large in-plane spin fluctuations and magnetic frustration upon doping might be at play [10]. Therefore, a doping-dependent study is essential for developing a deeper understanding of the relevance of cuprate physics to nickelate superconductivity.

To this end, we have examined the composition dependence of solid-solution $Nd_{1-x}Sr_xNiO_2$ infinite layer thin films for $0 \leq x \leq 0.25$. Given that cuprates provided an important context that stimulated the investigation of nickelates, we present our work in comparison to the canonical hole-doped $(La,Sr)_2CuO_4$ system [7]. We observe a superconducting dome for $0.125 < x < 0.25$, which is similar to hole-doped cuprates but half as wide. Unlike the cuprates, we observe weakly insulating behavior on either side of the dome. In addition, the lowest normal state resistivity is associated with the occurrence of superconductivity. Furthermore, the normal state Hall coefficient is close to zero and shows a continuous sign change both as a function of doping and temperature. We interpret this behavior in terms of a two-band picture with both electron and hole pockets, as suggested by electronic structure calculations.

Films of the infinite layer nickelate solid-solution $Nd_{1-x}Sr_xNiO_2$ were fabricated by pulsed-laser deposition of the perovskite precursor phase on (001) $SrTiO_3$ substrates, followed by topotactic reduction. We use the "high-fluence" conditions established in a recent study of the synthesis and microstructure of thin films of this compound [11]. These are the current optimal conditions for crystalline uniform films of up to ~ 10 nm in thickness, as studied here. The use of precise imaging conditions for laser ablation significantly improves the sample-to-sample reproducibility as compared to the initial report [4,11]. For all samples, nickelate films 8 – 10 nm in thickness were capped with a SrTiO$_3$ epitaxial layer, which serves as a structural supporting template during oxygen deintercalation using CaH$_2$. For samples with a SrTiO$_3$ cap layer of ~ 25 nm, the reduction temperature $T_r$ of 280 °C and reduction time $t_r$ of 4 – 6 h were used; for samples with a ~ 2 nm SrTiO$_3$ cap layer, $T_r$ and $t_r$ were 260 °C and 1 – 3 h, to achieve a complete transformation to the infinite layer phase. The annealing conditions were optimized as described previously [11]. No notable differences were observed for samples of the same composition with different capping layer thickness. High-angle annular dark-field scanning transmission electron microscopy (HAADF-STEM) images were acquired on samples prepared by focused ion beam, using an aberration-corrected FEI Titan Themis at 300 keV. Magnetotransport measurements were performed using Al wire-bonded contacts in a Hall bar geometry.

Figure 1(a) shows the *c*-axis lattice constant as a function of doping level ($x$ = 0, 0.1, 0.125, 0.15, 0.175, 0.2, 0.225, 0.25) for the resulting Nd$_{1-x}$Sr$_x$NiO$_2$ films at room temperature, averaged over 2 – 4 samples for each composition. The *c*-axis monotonically increases, approximately linearly, with increasing Sr doping from 3.28 Å at $x$ = 0 to 3.42 Å at $x$ = 0.25. This variation is consistent with the substitution of the larger Sr cation for Nd, and similar to the evolution observed in cuprates [7] although larger in scale, consistent with findings from band structure calculations [12]. It should be noted that for these films, in all cases the in-plane lattice constants are locked to the SrTiO$_3$ substrate, i.e., 3.91 Å, as verified by off-axis x-ray diffraction (XRD) [13].

Figure 1(b) displays the temperature-dependent in-plane resistivity $\rho_{xx}$ across the doping series. We note that unlike the *c*-axis lattice constants and the Hall coefficients discussed later, which are quite reproducible, $\rho_{xx}$ has nontrivial sample-to-sample variations (primarily in magnitude, although with similar temperature dependence). We speculate that these variations reflect the microstructure, in particular the distribution and density of extended defects based on vertical Ruddlesden-Popper-type faults that are commonly observed [14]. Figure 1(c) displays HAADF STEM images of regions exhibiting representative defects across the sample series (see Refs. [11,15] for details). While these do not significantly impact volume-sensitive measurements like XRD, the static transverse Hall field, or the overall systematic doping dependence, they do appear to contribute to scattering processes in $\rho_{xx}$. Therefore, in Fig. 1(b) we show $\rho_{xx}$ of a representative sample for each composition, and give an extended data set elsewhere [13].

The films with $x = 0.15, 0.175, 0.2$, and $0.225$ show varying $T_c$, while for $x = 0, 0.1, 0.125$, and $0.25$, a weakly insulating temperature dependence is observed down to the base temperature of a $He^3/He^4$ dilution refrigerator. These data indicate a superconducting dome that is similar to hole-doped cuprates [Fig. 2(a)], but approximately half as wide in density (closer to electron-doped cuprates [16]). While a small dip in $T_c$ for $x = 0.2$ is reproducibly observed, its significance is yet unclear. It is reminiscent of the "1/8" anomaly weakly visible in $(La,Sr)_2CuO_4$ [lower panel of Fig. 2(a)] and more strongly in $(La,Ba)_2CuO_4$ [17]. Furthermore, both $(La,Sr)_2NiO_4$ and $La_4Ni_3O_8$ show prominent 1/3 stripe order [18-21]. However, 1/5 is not a naturally expected density for ordering on a square lattice. Alternative explanations have also been proposed for related structure in superconducting nickelates [22].

Unlike the superconducting dome, which is qualitatively similar to the cuprates, the normal state transport properties evolve with notable differences. Upon hole doping, $\rho_{xx}$ in $(La,Sr)_2CuO_4$ drops by many orders of magnitude, consistent with the transition from insulator to metal with increasing carrier doping [7]. By contrast, the entire variation of the low-temperature normal state $\rho_{xx}$ in $Nd_{1-x}Sr_xNiO_2$ is approximately within an order of magnitude, even considering sample-to-sample fluctuations, and is at a minimum above the superconducting dome [Fig. 2(b)]. Interestingly, the normal state $\rho_{xx}(T = 20\ K)$ of the superconducting samples fall below the value (0.85 – 0.88 mΩ·cm across the doping series) that corresponds to the quantum sheet resistance ($h/e^2 \sim 26\ k\Omega \cdot ^{-1}$) per $NiO_2$ two-dimensional plane, as indicated in Fig. 2(b). Furthermore, an approximately linear $T$ dependence of $\rho_{xx}$ is observed across a wide temperature range above $T_c$, similar to that found in cuprates (known as the "strange metal" phase) [3] and other strongly correlated systems, suggesting a similar possible origin for $\rho_{xx}(T)$ despite different underlying electronic structure [23].

As for the noticeable upturn in $\rho_{xx}(T)$ for non-superconducting compositions ($x = 0, 0.1, 0.125$ and $0.25$), it is not simply identifiable as weak localization, in that the magnetoresistance (MR) of these samples is ubiquitously very small (discussed later and shown in Ref. [13]). The lack of strongly insulating behavior has already been broadly noted for the undoped compound [24], and has been suggested to reflect the onset of the Kondo effect [25-27]. Here we further observe that the "overdoped" regime does not appear to approach the Fermi liquid endpoint commonly understood to occur in the hole-doped cuprates [3]. Rather, the $x = 0.25$ sample exhibits as high $\rho_{xx}$ as for $x = 0$. To this end, increasing disorder induced by Sr substitution should be considered. Overall the relevance of the notion of "hole doping" is a question for (Nd,Sr)NiO$_2$.

To begin to address this point, we show the evolution of the normal state Hall effect across this series of samples [Fig. 3(a)]. In all cases, the Hall resistivity $\rho_{yx}$ shows no obvious deviation from linear dependence on magnetic field $\mu_0 H$ up to 8 T at all temperatures [13]. For the undoped sample, the room temperature Hall coefficient $R_H(300\ K)$ corresponds to ~ 0.08 electrons per formula unit, in the simplest approximation of $ne = 1/R_H$, where $n$ is the carrier density and $e$ is the electron charge. Upon Sr substitution, the magnitude of $R_H(300\ K)$ monotonically decreases, while maintaining negative sign. This would naively correspond to *increasing* electron density, completely counter to the expectation for hole doping. Furthermore, at high doping concentrations ($x = 0.2, 0.225, 0.25$), $R_H$ undergoes a smooth transition from negative at high temperatures to positive at low temperatures (below ~ 100 K). As a proxy for the variation in the low-temperature normal state Hall effect, the composition dependence of $R_H(20\ K)$ is given in Fig. 3(b). As $x$ increases from 0 to 0.25, $R_H(20\ K)$ monotonically increases from negative to positive values, crossing zero between $x = 0.175$ and 0.2.

Much of this behavior is in contrast to the hole-doped cuprates. In La$_{2-x}$Sr$_x$CuO$_4$, $R_H$ was found to be large and positive for the undoped case, and systematically varied as ~ $1/x$ with initial doping [7,28]. This began the debate on a small versus large Fermi surface, aspects of which continue to this day [3]. Only for extreme overdoping (~La$_{1.66}$Sr$_{0.34}$CuO$_4$) is similar behavior found, with a temperature dependent sign crossing at ~ 100 K [29], as in the $x = 0.25$ sample shown in Fig. 3(a). $R_H$ in this heavily hole-doped cuprate regime was attributed to a large Fermi surface composed of sections with different curvature [30]. Rather than the hole-doped cuprates, we note that in electron-doped RE$_{2-x}$Ce$_x$CuO$_4$ (RE, rare-earth), a similar zero crossing is observed in $R_H$ as functions of doping and temperature for densities associated with the superconducting dome [31,32].

The distinctive temperature and doping dependence of $R_H$ in the nickelates is clearly inconsistent with single-band hole doping. The simplest generalization we can consider is that of a two-band model with both electron-like and hole-like Fermi surfaces, depicted in the inset of Fig. 3(b). Such two-band analysis has been performed for strain- and temperature-dependent $R_H$ in thin film NdNiO$_3$ for example [33]. For the purposes of discussion, we quote the magnetotransport relations of this model [34]:

$$\rho_{yx} = \frac{1}{e} \frac{(n_h \mu_h^2 - n_e \mu_e^2) + \mu_h^2 \mu_e^2 B^2 (n_h - n_e)}{(n_h \mu_h + n_e \mu_e)^2 + \mu_h^2 \mu_e^2 B^2 (n_h - n_e)^2} B$$

$$\rho_{xx} = \frac{1}{e} \frac{(n_h \mu_h + n_e \mu_e) + (n_e \mu_e \mu_h^2 + n_h \mu_h \mu_e^2) B^2}{(n_h \mu_h + n_e \mu_e)^2 + \mu_h^2 \mu_e^2 B^2 (n_h - n_e)^2}$$

where $n_e$ ($n_h$) is the electron (hole) density and $\mu_e$ ($\mu_h$) is the electron (hole) mobility. In the low-field limit, $R_H$ reflects the difference between $n_e$ and $n_h$, as weighted by their squared mobilities. Therefore, this two-band model can fully capture the evolution of $R_H$ that we observe: with increasing hole doping via Sr, a predominantly electron-like Hall effect transitions to a predominantly hole-like Hall effect, following the decrease in Fermi level. The temperature dependence reflects the different dependencies of $\mu_e$ and $\mu_h$.

This two-band picture is well corroborated by many recent electronic structure calculations. The general consensus of density functional theory for $NdNiO_2$ is the presence of a large hole pocket with Ni $3d_{x^2-y^2}$ character, and two electron pockets with Nd $5d_{xy}$ and $5d_{3z^2-r^2}$ character [6,12,26,35-47]. With increasing Coulomb interaction on the Ni site, the $3d_{x^2-y^2}$ band opens a Mott gap forming an upper and lower Hubbard band. Unfortunately, while our experiments are quite consistent with these studies, we cannot quantitatively extract estimates for the specific densities and mobilities, as the linear Hall response is highly under-constrained [13]. Furthermore, the zero-field resistivity or magnetoresistance we observe [13] does not provide useful further constraints. Overall, the normal state MR is extremely small (below 1% at 8 T for all compositions and temperatures), and negative, whereas the simple two-band model would predict a positive definite response. Furthermore, the sample-to-sample variations in $\rho_{xx}$ noted earlier affect the interpretation of the MR. Further refinements in the sample quality, and the understanding of the relevant transport scattering mechanisms, are clearly needed.

In summary, we find a superconducting dome in the phase diagram of solid solution $Nd_{1-x}Sr_xNiO_2$ infinite layer thin films for $0.125 < x < 0.25$. From our observations we can phenomenologically associate the occurrence of superconductivity with a minimum in the normal state $\rho_{xx}$, and a normal state $R_H$ close to zero but of either sign, suggestive of a two-band picture. For scenarios where pairing is dominated by the $d_{x^2-y^2}$ band [22,36-38,48-51], this may simply correspond to the appropriate hole doping level required for superconductivity in the lower Hubbard band, resulting in similar superconducting domes in the nickelates [47] and cuprates. For scenarios that additionally consider Kondo coupling between the band electrons derived from Nd $5d$ states and the Ni spin-1/2 lattice, the occurrence of superconductivity corresponds to increasing exhaustion of the screening electron density [12,25-27].

In terms of the normal state $\rho_{xx}$, it has already been noted that NdNiO$_2$ is not a proper insulator, and rather that it can be considered a self-doped Mott insulator by virtue of the two-band degree of freedom [6,25]. Here we additionally find that in the overdoped regime beyond superconductivity, a normal state $\rho_{xx}$ is observed which is as high as the undoped compound. In cuprates, the relevant states for conduction reside in the CuO$_2$ plane, and the interlaced rare-earth spacer layers usually provide a simple "charge reservoir" controlling the carrier density in the CuO$_2$-derived band(s). With increasing doping away from the Mott insulator, the ionized dopants are increasingly well screened [52], and a metal is observed beyond the superconducting dome. By contrast, the Nd layer in NdNiO$_2$ not only acts as a charge reservoir, it also hosts a relevant electron band. Here, increasing Sr substitution introduces increasing disorder, which may play a role in the very different behavior in the overdoped regime.

It should be noted that despite the systematic evolution of the lattice parameters and Hall effect, there may also be systematic contributions from materials imperfections and defects that monotonically vary with Sr substitution. This is particularly relevant here given the unconventional Ni oxidation states that are traversed in synthesizing this metastable compound, which warrants further investigation. Nevertheless, it is already clear that the normal state of the nickelates is qualitatively different from the cuprates. To the extent that a general consensus is that a microscopic understanding of cuprate superconductivity requires first an understanding of the normal state [3], the distinct normal state properties of the nickelates presents a similar challenge.

We thank Malcolm R. Beasley, Richard L. Greene, Nigel E. Hussey, Wei-Sheng Lee, Brian Moritz, George A. Sawatzky, and Hidenori Takagi for discussions. The work at SLAC and Stanford was supported by the U.S. Department of Energy, Office of Basic Energy Sciences, Division of Materials Sciences and Engineering, under Contract No. DE-AC02-76SF00515, and the Gordon and Betty Moore Foundation's Emergent Phenomena in Quantum Systems Initiative through Grant No. GBMF4415 (dilution refrigerator experiments and synthesis equipment). M.O. acknowledges partial financial support from the Takenaka Scholarship Foundation. The work at Cornell was support by the Department of Defense, Air Force Office of Scientific Research (No. FA 9550-16-1-0305) and the Packard Foundation. This work made use of a Helios FIB and a Titan Themis 300 supported by NSF (DMR-1539918, MRI-1429155), the Cornell Center for Materials Research Shared Facilities which are supported through the NSF MRSEC program (DMR-1719875) and the Kavli Institute at Cornell.

*Note added*. Recently, we became aware of another study [53] on the thin-film Nd$_{1-x}$Sr$_x$NiO$_2$ system, which found a similar superconducting dome structure and normal-state transport behavior, albeit with somewhat different magnitudes.

*D.L. and B.Y.W. contribute equally to this work
†denverli@stanford.edu
‡bwang87@stanford.edu
§hyhwang@stanford.edu

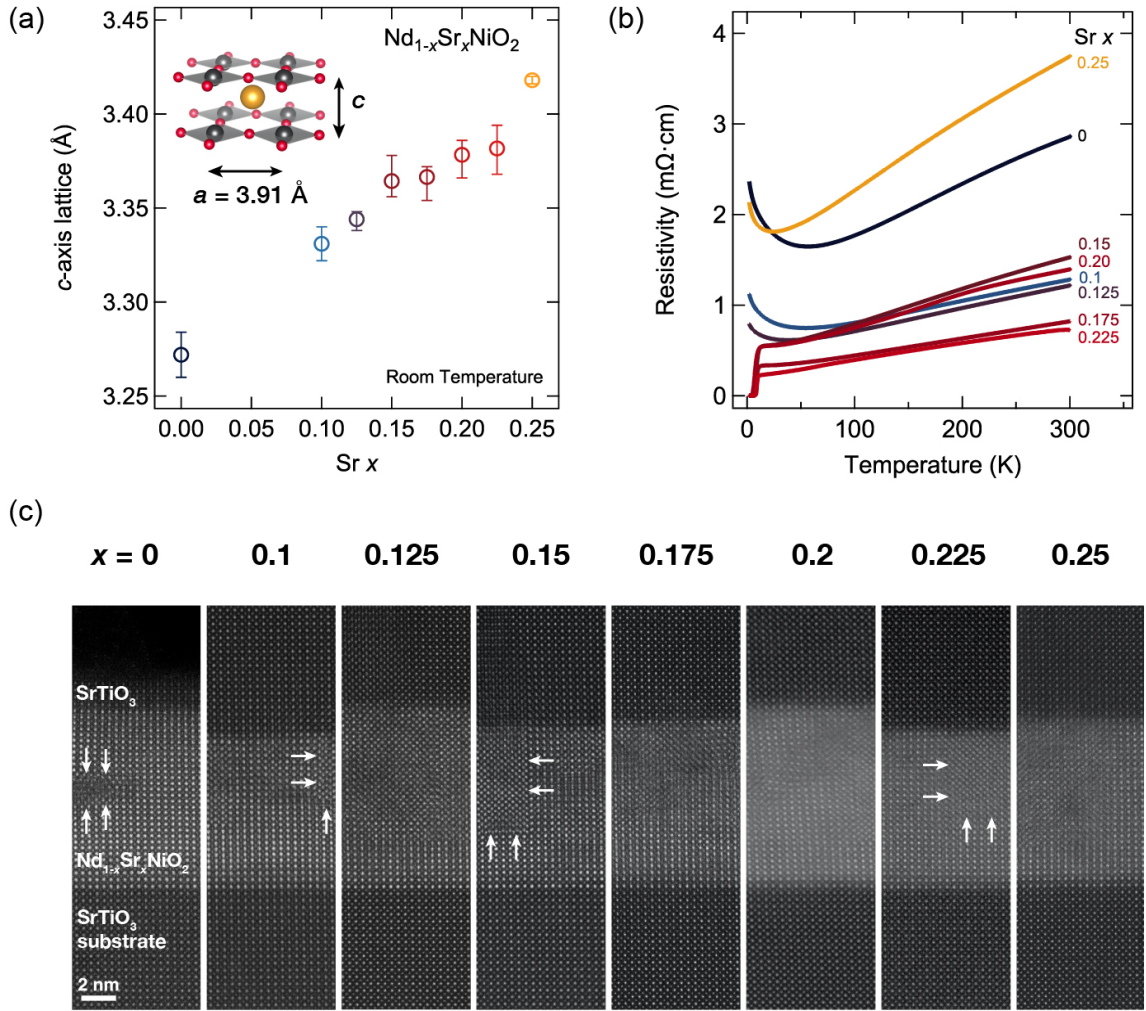

FIG.1 (color online). Structural and transport properties as a function of $x$ in $Nd_{1-x}Sr_xNiO_2$ thin films. (a) Room temperature $c$-axis lattice constants as a function of Sr substitution. The circles represent the average value across multiple samples of the same $x$. Error bars indicate the extremal values. (b) Temperature dependent resistivity (300 – 2 K) measured for representative samples with different doping levels $x$ = 0, 0.1, 0.125, 0.15, 0.175, 0.2, 0.225, and 0.25 (see Ref. [13] for an extended dataset). (c) HAADF-STEM images across the sample series exhibiting representative defects, primarily derived from Ruddlesden-Popper-type regions (indicated by the white arrows). For $x$ = 0, the $SrTiO_3$ cap was 2 nm; for the rest, the cap was 25 nm.

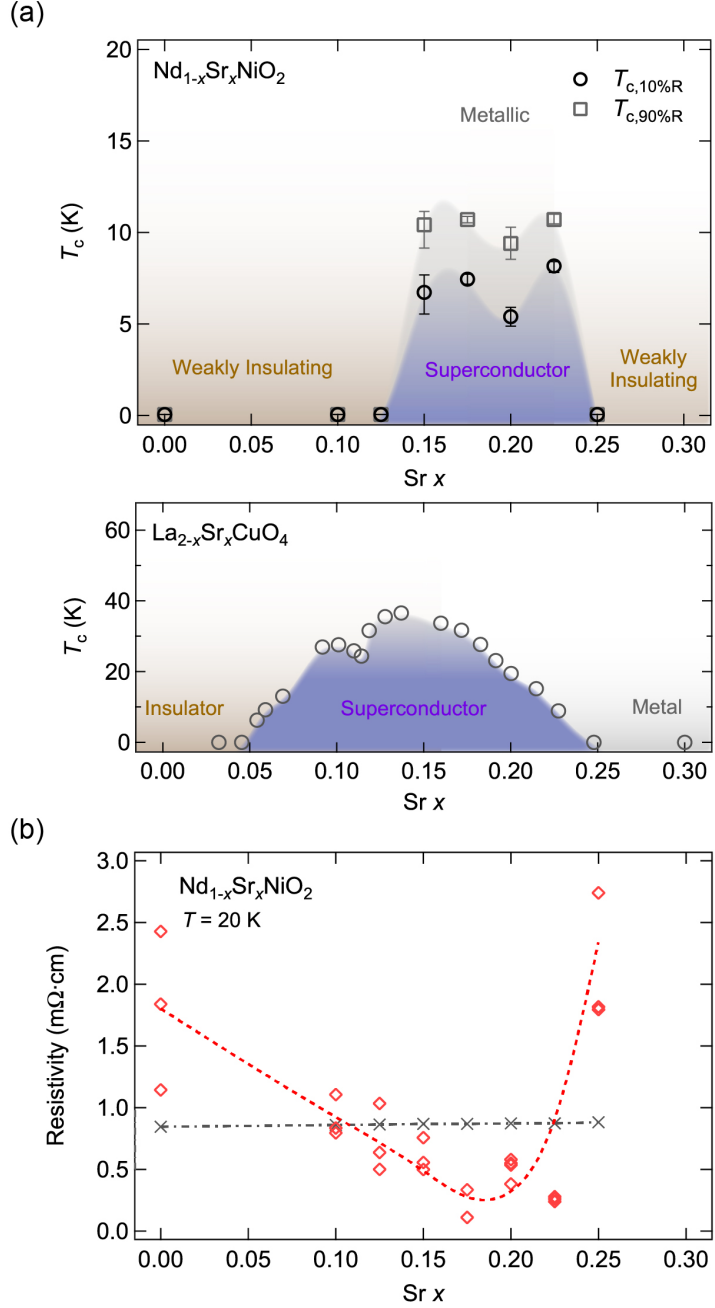

FIG.2 (color online). (a) Top: Phase diagram of $Nd_{1-x}Sr_xNiO_2$. Open circles (squares) represent $T_{c,10\%R}$ ($T_{c,90\%R}$), as defined to be the temperatures at which the resistivity is 10 % (90 %) of the resistivity value at 20 K. The symbols denote the average value across multiple samples of the same $x$. Error bars indicate the extremal values. For $x$ = 0, 0.1, 0.125, and 0.25, no sign of superconductivity is observed down to < 50 mK. Bottom: Phase diagram of $La_{2-x}Sr_xCuO_4$ with $T_{c,50\%R}$ values adapted from Ref. [7]. (b) Resistivity at 20 K measured for all samples as a function of $x$. The dashed line is a guide to the eye. The crosses connected by a horizontal dot-dashed line indicate the resistivity values corresponding to a resistance quantum per $NiO_2$ plane, which are calculated using the $x$ dependent $c$-axis lattice constants.

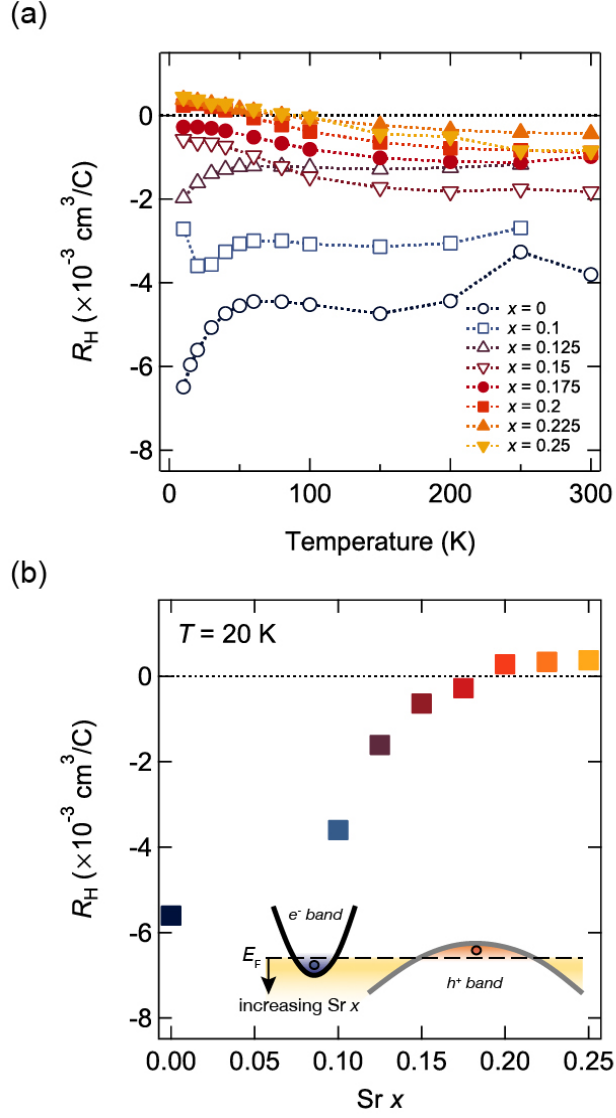

FIG.3 (color online). Normal state Hall coefficient $R_H$ for $Nd_{1-x}Sr_xNiO_2$ films. (a) Temperature-dependent $R_H$ from 300 K down to 10 K for various $x$ (0, 0.1, 0.125, 0.15, 0.175, 0.2, 0.225, and 0.25). (b) $R_H$ as a function of $x$ at 20 K, crossing zero between $x = 0.175$ and 0.2. The inset displays a two-band schematic (electron pocket, $e^-$ band; hole pocket, $h^+$ band). The arrow indicates the variation of the Fermi level $E_F$ with increasing $x$.

**Supplemental Material** *for*

**Superconducting dome in Nd$_{1-x}$Sr$_x$NiO$_2$ infinite layer films**

Danfeng Li[1,2], Bai Yang Wang[1,3], Kyuho Lee[1,3], Shannon P. Harvey[1,2], Motoki Osada[1,4], Berit H. Goodge[5], Lena F. Kourkoutis[5,6] & Harold Y. Hwang[1,2]

[1]*Stanford Institute for Materials and Energy Sciences, SLAC National Accelerator Laboratory, Menlo Park, CA 94025, USA*
[2]*Department of Applied Physics, Stanford University, Stanford, CA 94305, USA*
[3]*Department of Physics, Stanford University, Stanford, CA 94305, USA*
[4]*Department of Materials Science and Engineering, Stanford University, Stanford, CA 94305, USA*
[5]*School of Applied and Engineering Physics, Cornell University, Ithaca, New York 14853, USA*
[6]*Kavli Institute at Cornell for Nanoscale Science, Ithaca, New York 14853, USA*

This Supplemental Material includes the following datasets: (1) X-ray reciprocal space maps for $Nd_{1-x}Sr_xNiO_2$ samples with $x = 0$ and $x = 0.25$ (Fig. S1); (2) Extended dataset of $\rho_{xx}(T)$ for multiple samples measured across different Sr concentrations (Fig. S2); (3) Anti-symmetrized Hall resistivity $\rho_{yx}(H)$ dataset (Fig. S3) for two representative samples ($x = 0, 0.225$); (4) Magnetoresistance of $Nd_{1-x}Sr_xNiO_2$ films at various temperatures across different $x$ (Fig. S4).

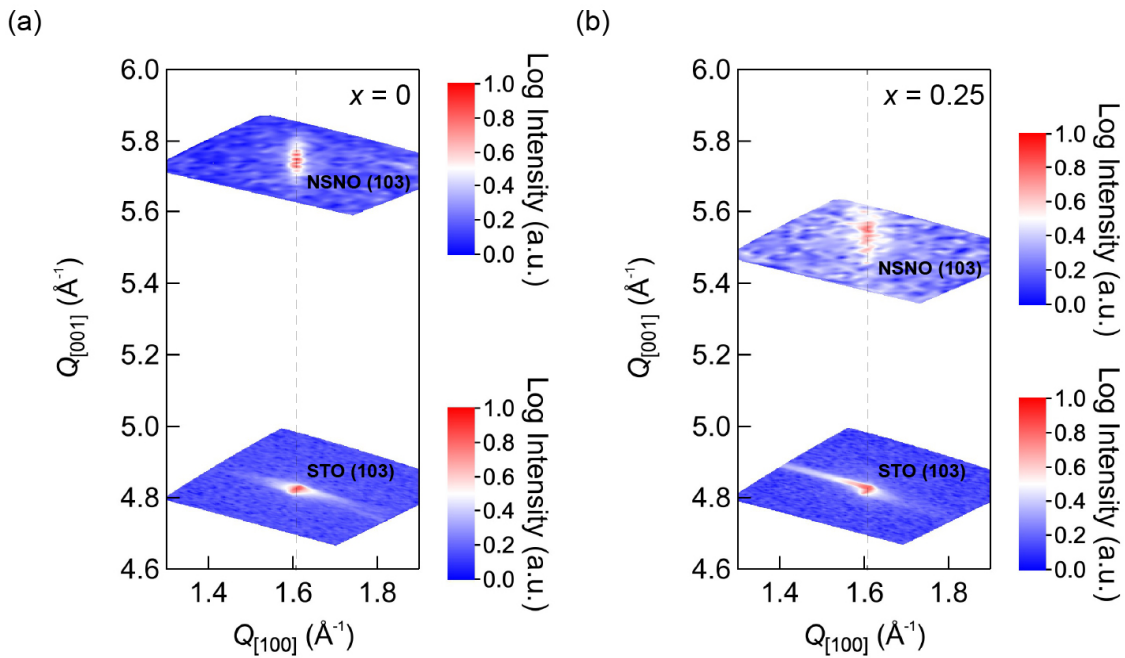

FIG. S1. X-ray reciprocal spacing maps around (103) diffraction for $Nd_{1-x}Sr_xNiO_2$ films on $SrTiO_3$ (001) substrates at room temperature. (a) $x = 0$; (b) $x = 0.25$. STO, $SrTiO_3$; NSNO, $Nd_{1-x}Sr_xNiO_2$; a.u., arbitrary units. The vertical dashed lines indicate that the films are fully strained to the substrates.

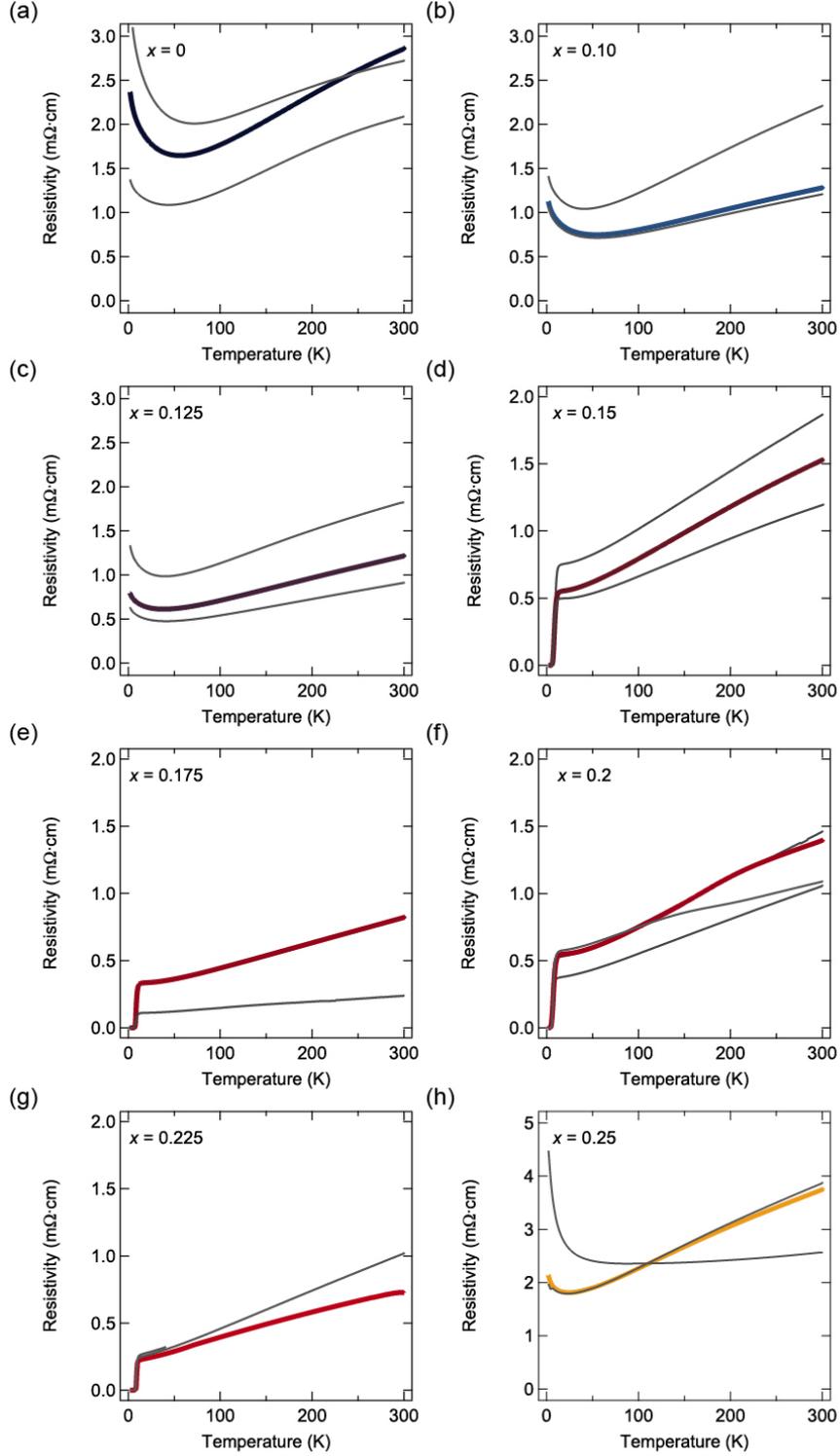

FIG. S2. Resistivity as a function of temperature (300 K to 2 K) of multiple samples for different $x$. (a) $x = 0$; (b) $x = 0.1$; (c) $x = 0.125$; (d) $x = 0.15$; (e) $x = 0.175$; (f) $x = 0.2$; (g) $x = 0.225$; (h) $x = 0.25$. The thick colored curves are the ones presented in Fig. 1(b) using the same color code. The data for other samples are displayed in grey.

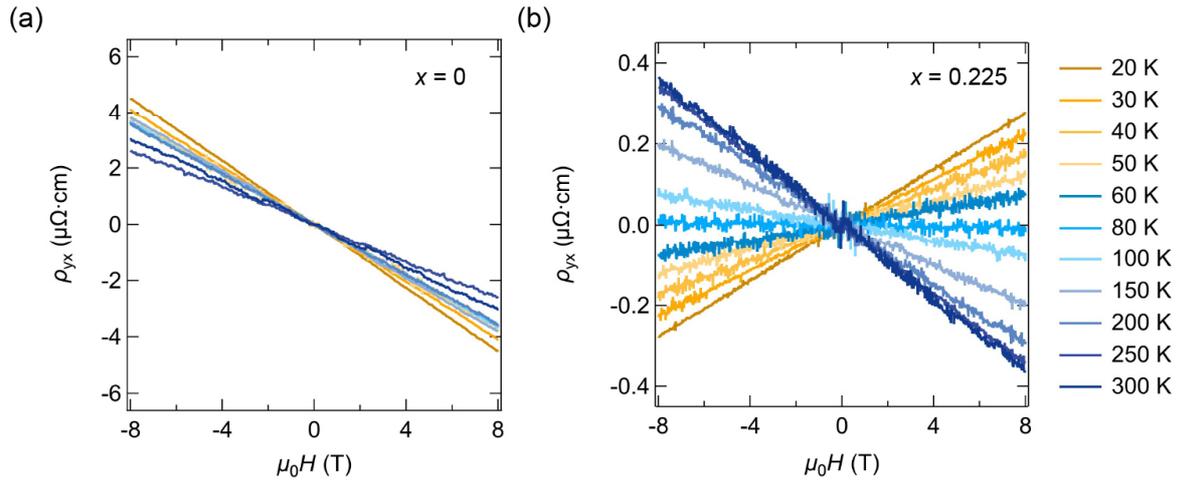

FIG. S3. Hall resistivity ($\rho_{yx}$) as a function of magnetic field ($\mu_0 H$) measured at different temperatures for (a) $x = 0$ and (b) $x = 0.225$. The temperatures are listed on the right.

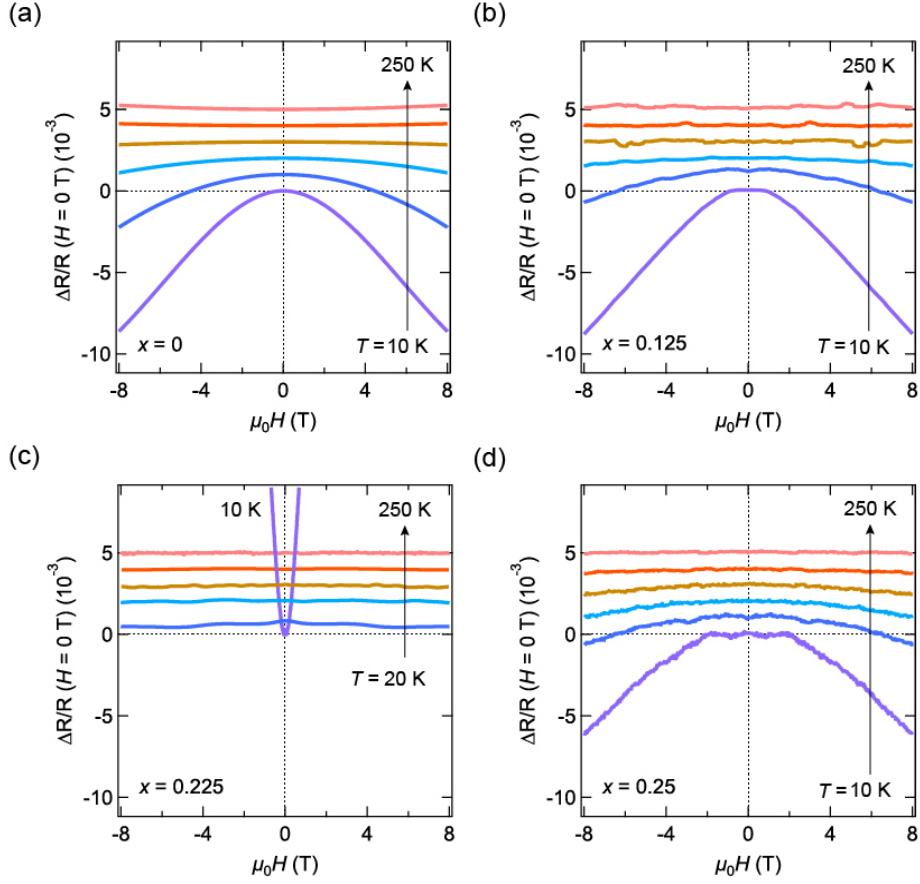

FIG. S4. Magnetoresistance of $Nd_{1-x}Sr_xNiO_2$ films at various temperatures across different $x$: (a) $x = 0$; (b) $x = 0.125$; (c) $x = 0.225$; (d) $x = 0.25$. The magnetic field is applied parallel to the $c$-axis (out-of-plane). The data shown are at 10 K, 20 K, 40 K, 80 K, 150 K, and 250 K. The data are normalized by their zero field resistance values and vertically shifted in increments of 0.001 with respect to the 10 K data, for clarity.